\documentclass[preprint,showpacs,preprintnumbers,amsmath,amssymb]{revtex4}

\usepackage{graphicx}
\usepackage{dcolumn}
\usepackage{bm}

\begin{document}

\preprint{APS/123-QED}

\title{Quantum computing with single electron bubbles in helium}

\author{Weijun Yao}
\email{Weijun_Yao@brown.edu}
\affiliation{Department of Physics, Brown University, Providence,
Rhode Island 02912}


\begin{abstract}
An electron inside liquid helium forms a bubble of 17~\AA~in radius.
In an external magnetic field,
the two-level system of a spin 1/2 electron is ideal
for the implementation of a qubit for quantum computing.
The electron spin is well isolated from other thermal reservoirs
so that the qubit should have very long coherence time.
By confining a chain of single electron bubbles in a linear RF quadrupole trap,
a multi-bit quantum register can be implemented.
All spins in the register can be initialized to the ground state either by establishing
thermal equilibrium at a temperature around 0.1 K and at a magnetic field of 1~T or by
sorting the bubbles to be loaded into the trap with magnetic separation.
Schemes are designed to address individual spins and to do two-qubit CNOT operations between the neighboring spins.
The final readout can be carried out through a measurement similar to the Stern-Gerlach experiment.

\end{abstract}

\pacs{03.67.Lx, 67.40.Yv, 47.55.Dz}
\maketitle

%

\section{Introduction}

Recent research indicates that quantum computing can effectively solve
some problems that are considered intractable on classical Turing machines.
A sample of such problems is the factorization of large composite numbers into primes,
a problem which is the basis of the security of many classical key cryptosystems \cite{Nielsen00}.
The task of designing a quantum computer is equivalent to finding a physical system of
quantum bits (qubits) and quantum gates between the bits.
Such a physical system has to fulfill several basic criteria~\cite{DiVincenzo00}.
First, one needs a system with two eigenstates for each qubit
which can be identified as the two logic states $|0\rangle$ and $|1\rangle$.
One should be able to perform one-qubit operations
on an arbitrary qubit without affecting the other qubits.
The operations should be able to create a state which is the superposition of the two
eigenstates.
Second, the qubits must interact with each other to provide a set of universal logic gates.
It has been shown that any operation can be
decomposed into controlled-not gates (CNOT) of two qubits between the nearest neighbor
and rotations on individual qubits \cite{DiVincenzo95}.
Third, one must be able to initialize the qubits to a known state at the beginning
of each operation sequence since the results of computation generally depend on the input states.
Fourth, one should be able to extract the results from the qubits by some measurements.
Fifth, the quantum mechanical feature of the
qubits also require that they are well isolated from environment so that the
coherence time is long compared with an average logic gate duration
such that many logic gates can be implemented within the coherence
time.  This means for each qubit, the quantum state which is a superposition
of two eigenstates, should sustain itself long enough before relaxing to the state determined
by the environment.
Furthermore, a useful register for a quantum computer needs at least
$10^3$ qubits. Scalability of proposed quantum computer
architecture to many qubits is of central importance \cite{Cirac95}.

Meeting all these criteria simultaneously pose a significant
experimental challenge.
For example, how can one gain access to a
quantum system and at the same time keep the coherence time long?
So far only a few systems have been identified as potential
viable quantum computer models.
Atoms \cite{Monroe02} and ions in traps \cite{Kielpinski02},
cavity quantum electrodynamic system \cite{Raimond01},
nuclei in complex molecules \cite{Cory97},
quantum dots and other solid state devices \cite{Maklin01}
and electrons confined on top of liquid helium \cite{Platzman02}
have been studied as possible implementations.
In this paper, we propose a system where individual
electrons merged inside superfluid liquid helium are used as qubits.
The paper is arranged in such a way that first we will describe
some properties of single electrons inside liquid helium,
then we will discuss how to confine a chain of such electrons inside a trap.
In the remainder of the paper, we will focus on how such a system can fulfill
the criteria for a quantum register.

\section{Single Electron Bubbles}

When an energetic electron enters liquid helium, it dissipates its
energy quickly through producing excited state helium atoms and
quasi-particles like rotons and phonons. After the electron comes to rest,
it will expel the surrounding helium atoms and produce a
spherical cavity of approximately 17 \AA~in radius (see below) which is almost free of helium atoms.
The properties of such single electron
bubbles have been studied extensively \cite{Maris03}.
Here we just focus briefly on some features which are relevant to the implementation of qubits.

The total energy of such an electron bubble can be estimated by
\begin{equation}
E=E_{\rm el} + \gamma A +PV
\end{equation}
where $E_{\rm el}$ is the energy of the electron,
$\gamma$ is the surface energy of liquid helium,
$A$ is the surface area of the bubble,
$P$ is the pressure inside the liquid and $V$ is the volume of the bubble.
In the first order approximation,
we can consider that the electron is confined inside a spherical square well
and the wave function of the electron goes to zero at the wall of the bubble.
The ground state (1s) energy of the electron $E_{\rm el}$ can be written as
\begin{equation}
E_{el}=\frac{h^2}{8mR^2},
\label{electron_energy}
\end{equation}
where $m$ is the mass of the electron, $R$ is the bubble radius.
The radius of the bubble thus can be determined by minimizing the total energy $E$.
If assuming the pressure $P$ is zero,
this gives
\begin{equation}
R=\left(\frac{h^2}{32\pi m \gamma} \right)^{1/4}
\label{bubble_radius}
\end{equation}
If the value of $0.37~$erg/cm$^2$ is taken for the surface tension of liquid helium-4~\cite{Vicente02},
Eq.~\ref{bubble_radius} gives $R=19$~\AA.
Several factors such as the finite thickness of the liquid vapor interface,
the penetration of the electron wave function into the liquid helium can give some corrections to the
above calculated $R$ value \cite{Maris00}.
Comprehensive reviews of electrons in helium can be found in reference \cite{Fetter74}.

From the size of an electron bubble,
the effective mass of such an object can be estimated to be equal to
the amount of liquid helium expelled.
Experimentally, the effective mass was
measured, which gives the value of $243$~m$_{\rm He}$,
m$_{\rm He}$ is the mass of one helium atom~\cite{Poitrenaud72, poitrenaud74}.
In the next section, this should be the value to be used to calculate the effective trapping potential
of a linear quadrupole trap.

The energy spectra of single electron bubbles were investigated both theoretically
and experimentally \cite{Maris03, Grimes90}.
Here we are interested in the laser induced fluorescence when the electron is excited from 1s to 1p or
2p state.
The photons emitted from the bubbles can be used to indicate their locations.
We will discuss how to use such information in reading out the qubits.
The transition energy of 1s$\rightarrow$1p is 0.11 eV and 1s$\rightarrow$2p is 0.52 eV.

The two-level system of an electron bubble is due to two eigenstates
of the spin 1/2 electron in a magnetic field.
The $g$ value of such an electron spin was measured and it is equal to the $g$ value of
a free electron within 10~ppm \cite{Reichert74}.
For the convenience of later discussions,
we define the lower energy state in which the spin is parallel to the magnet field, $|0\rangle$,
and the higher energy state in which it is anti-parallel to the field, $|1\rangle$.
These two states are the two logic states of a quantum qubit.
Similarly, for two qubits (two single electron bubbles),
we use the symbols $|00\rangle$, $|10\rangle$, $|01\rangle$ and $|11\rangle$
to represent the combined states.
Here we emphasize that the multi-bits system are
multiple single electron bubbles with one electron in each bubble in contrast to
a multiple electron bubble in which many electrons are confined in a single bubble.

An important requirement of the implementation of quantum computing is that the
coherence time of an individual qubit should be much longer than the duration of a quantum operation.
The electron in the helium void has very weak interactions with its surroundings.
The source of decoherence is the spin-lattice relaxation.
Some theoretical calculation predicts that the spin-lattice relaxation time $T_1$
could be in the order of $10^5$ seconds~\cite{Huang71}.
So far there was only one experimental investigation on $T_1$ through an ESR measurement,
which showed that $T_1$ was longer than 100 ms \cite{Reichert74}.
It will be useful to extend the investigations.
Liquid helium is such a pure system that away from the walls of its container
there are no impurities (except the isotope helium 3) that can introduce spin-lattice relaxation.
It is worth mentioning that helium~3 which has nuclear spin 1/2 can
influence $T_1$~\cite{Reichert83}.
There is around 0.1-0.2 ppm of helium~3 in helium~4.
But it is not clear if the trace amount of helium~3 will have any significant effect on $T_1$.

\section{Trapping single electron bubbles}

The technique of using radio frequency trap (Paul trap) to confine
ions in vacuum have been developed in the research of
precision spectroscopy and standards,
laser cooling and recently in the implementation of quantum computing \cite{Paul90, Brewer92}.
A single electron bubble is not much different from a very heavy ion.
It has a mass of 243 times the mass of a helium atom and a charge of an electron.
It is inside a dissipative environment if the temperature is not too low.
Since trapping single or multi-electron bubbles can be a useful technique in the study such as
bubble fission \cite{Maris00} and degenerate two dimensional electron systems \cite{Silvera05},
we will devote a separate paper to describe various trapping schemes.
Here we describe briefly a linear quadrupole trap to confine a chain of single electron bubbles as a quantum register.

\begin{figure}[htbp]
\centering \includegraphics[scale=0.5]{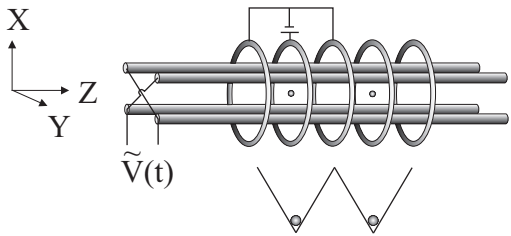}
\caption{A quadrupole trap. The periodical static potential is produced by the
rings to confine the bubbles in the z direction. In some situations as described in the text,
several neighboring rings' potential should be tuned individually.}
\label{fig_trap}
\end{figure}

The device consists of four parallel rods as shown in Fig.~\ref{fig_trap}.
Typically, each rod has a circular cross section.
A time-varying RF potential $V_0\cos\Omega t$ is applied to the two opposing rods.
The remaining two rods are held at RF ground.
The time-varying potential near the axis can be approximated by
\begin{equation}
V=\frac{V_0}{2}\left(1+\frac{x^2-y^2}{r^2}\right) \cos \Omega t,
\end{equation}
where $r$ is the distance from the axis to the surface of the electrodes.
For sufficiently high driving frequency $\Omega$,
a particle with mass $m$ and charge $q$ moves in the effective pseudo-potential
\begin{equation}
\Phi=\frac{qV_0^2}{4m\Omega^2r^4}(x^2+y^2)=\frac{m}{2q}w_r^2(x^2+y^2),
\label{eq_trap}
\end{equation}
where
\begin{equation}
w_r=\frac{qV_0}{\sqrt{2}m\Omega r^2}
\end{equation}
To make a stable parabolic trap as described by Eq.~\ref{eq_trap},
the RF driving frequency $\Omega$ has to satisfy
\begin{equation}
\frac{2qV}{mr^2\Omega^2} \le 0.92
\end{equation}
If $r \sim 1$~mm and $V \sim 1$~Volt,
this will give a minimum value of 74~kHz.
From Eq.~\ref{eq_trap}, one can see for moderate voltages and frequencies,
the trap is very steep.
After a bubble is trapped inside the trap,
its amplitude of thermal motion in x-y plane is very small.
So all the bubbles form a one dimensional chain
and they can not interchange their positions.

The confinement in the $z$ direction is achieved by a static electric field
produced by the rings at different potential
which is schematically shown in Fig.~\ref{fig_trap}.

\section{Single Qubit and Multiple Qubits}

The chain of electron bubbles in the Fig. \ref{fig_trap}
is similar to a big artificial one-dimensional molecule.
A magnetic field is applied along the $z$ direction.
For a single qubit operation,
a RF pulse at the Larmor frequency can be applied to rotate the spin.
By introducing a magnetic field gradient along the z axis,
$B=\alpha z$,
each spin experiences a different field and has a unique Larmor frequency.
This way, all the spins can be addressed individually.
The situation discussed here is very similar to the NMR quantum
computing where the quantum computer consists of several
individual atoms with spin $1/2$ nuclei~\cite{Cory97}.
Different nuclei experience different total (external plus internal)
magnetic field thus have different Larmor frequency.
The interaction between two neighboring spins is via magnetic coupling.

At the beginning of each computation sequence,
the quantum register,
here is the chain of spins,
has to be set to a known initial state, usually to its ground state.
This can be achieved by establishing thermal equilibrium between the spin ensemble and the external
thermal reservoir of the liquid helium.
At 4.2 K and a field of 1 T, the probability of finding a spin in the ground state is $58\%$;
if $B/T$ is increased by a factor of 20 and at 0.2 K, $99.98\%$.
As mentioned above, the coupling between the electron spin with the surrounding helium
is very weak.
On one hand, the qubits benefit from such weak interaction and have very long coherence time.
On the other hand, this makes thermalization a very slow procedure.
The time needed can be estimated by
\begin{equation}
\frac{dp}{dt}=-\frac{p-p_{\rm eq}}{T_1},
\end{equation}
where, $p$ is the polarization, $p_{\rm eq}$ is the equilibrium polarization,
and $T_1$ is the spin-lattice relaxation time.
So, a waiting period of several $T_1$ is needed to let the system being
settled to the ground state before each operation sequence.
This is certainly not practical if $T_1$ is in the order of $10^5$~s.
Another possible solution to this issue is that during the loading procedure of the bubble chain,
one can sort the bubbles according to their spin orientations and let only the bubbles
with desired spins enter the trap. Since $|0\rangle$ bubbles tend to
move toward high field regime and $|1\rangle$ bubbles move the other way, the separation can
be accomplished by a magnetic field gradient.

After a sequence of operations, we need a mechanism to read out the result of each qubit.
The following is a scheme which is similar to the Stern Gerlach experiment.
By lowering the electric potential barrier on the end of the chain,
the bubble at the end will start to drift. Owing to the gradient of the magnetic field,
the bubble will experience a force whose direction depends on the spin orientation.
The force,
\begin{equation}
F= -g \mu_B s \frac{\partial B}{\partial z}=-g \mu_B s \alpha,
\end{equation}
where $\mu_B$ is the Bohr magneton, $s$ is the spin, and $g$ is the g-value of the spin.
There will be two different drift velocities corresponding to the two spin orientations.
After a period of time,
we can take a snap picture by shining a laser bean toward the two possible locations of the bubble and
we should see the bubble at only one place. This information will tell the spin orientation.
And repeating the same procedure for all the bubbles, one can get a measurement of the whole register.

The scalability of the scheme depends on how large the magnetic gradient can be applied and how well
the RF instruments can separate the frequencies to address the individual spins.
The higher the gradient, the denser the bubbles in the trap.

\section{Two qubit controlled-not operation}

The two qubit controlled-not operation is defined as
\begin{eqnarray}
| 00 \rangle \rightarrow |00 \rangle~~ | 10 \rangle \rightarrow |10 \rangle \\
| 01 \rangle \rightarrow |11 \rangle~~ | 11 \rangle \rightarrow |01 \rangle
\label{eq_two_bit}
\end{eqnarray}
where the second bit is the control bit.
The CNOT operation in NMR quantum computing was achieved through
the spin-spin interaction. The magnetic resonant frequency of a particular spin
depends on the orientations of its neighboring spins.
Such interaction is strong only within interatomic distance in a molecule.
To bring a pair of electron bubbles to such a distance,
the external electric field has to be exceedingly large.

\begin{figure}[htbp]
\centering \includegraphics[scale=0.38]{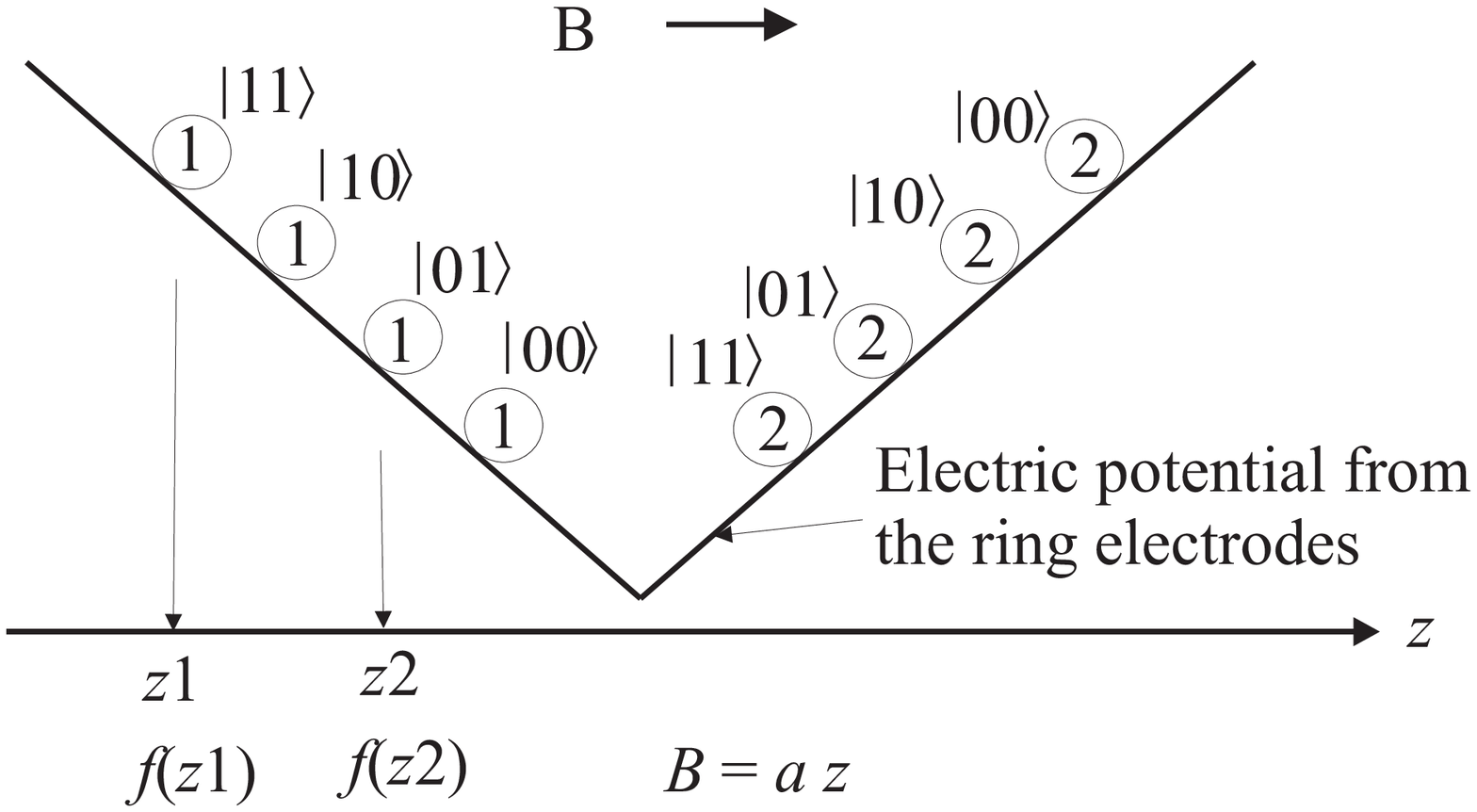}
\caption{Two-bit controlled not operation.
For $|00\rangle$ and $|11\rangle$, the centers of the pair shift
due to the magnetic field gradient but the distances between the bubbles keep the same.
While for $|10\rangle$ and $|01\rangle$,
the centers keep the same but the distances change.}
\label{fig_two_bit}
\end{figure}

The following scheme is designed to address the two qubit operation by making use of the magnetic
field gradient.
By lowering the electric potential barrier that separates the neighboring bubbles,
the two bubbles can be brought closer inside the same trap.
Due to the Coulomb repulsion,
the two bubbles will be separated by a distance which can be controlled by the strength of
the static potential. Besides the Coulomb interaction,
the spins will also experience a force due to the magnetic field gradient.
As shown in Fig. \ref{fig_two_bit},
there exist four sets of locations for the pair. And for bubble 1,
$z1$ and $z2$ are the locations for states $|11\rangle$ and $|01\rangle$, respectively.
As indicated also in the figure, frequency $f(z1)$ will cause $|11 \rangle \rightarrow |01 \rangle$
and $f(z2)$ will cause $| 01 \rangle \rightarrow |11 \rangle$  transition.
By applying a $\pi$ pulse which contains the two frequencies simultaneously,
we can execute the operation specified in Eq.~\ref{eq_two_bit} without altering the other states.
This is the two bit controlled-not.
After the operation,
the two bubbles can be re-separated by raising barrier between them again.

In the above discussion, we assume that a bubble's motion follows Stokes law that
the velocity is proportional to the force applied on it.  When the force
is reduced to zero, the bubble stops moving. This condition is satisfied at temperatures
above 0.5 K in pure helium-4 when there are still enough photons and rotons to dissipate the
energy of the bubbles.
At lower temperatures, some other dissipation mechanism has to be introduced since the density of those
quasi-particles becomes very low. The issue can be solved by simply adding several tens of ppm of helium-3
into the system which will significantly increase the dissipation since the helium-3 inside will not become
superfluid at the lowest temperature ever achieved.

In this paper, we discussed the possibility of the implementation of quantum computing with a chain of single
electron bubbles inside liquid helium.
The method combines the schemes of trapped ions and NMR quantum computation.
Such a system should be able to scale up to $10^2$ bits without major technical obstacles.

Finally, we would like to mention that similar schemes may also work for positive ions with
one unpaired electron.
For example, the alkaline earth metal ion Ba$^+$, Mg$^+$ and Sr$^+$.
Some spectroscopic studies have been carried out recently on those specimens immersed inside liquid
helium \cite{Tabbert95}.
One possible advantage of those systems is that there may exist some optical transitions
that can be used to manipulate and measure the spin orientation of individual ions through laser pulses.
Positive helium ions can also be made.
A difference between positive ions and
electrons inside helium is that positive ions attract the neighboring helium atoms and form solid
balls instead of cavities. Some further work is still needed in the investigation of those
systems in the possible applications in quantum computing.

The author would like to thank Dr. S. Balibar, Prof. G. Seidel and Prof. I. Silvera
for very helpful discussions and suggestions.

\bibliography{bib/qubit}
\end{document}